# 2MASS Identifications for Galactic OB Stars

B. Cameron Reed

Department of Physics

Alma College

Alma, MI 48801

e-mail: reed@alma.edu

ph: (989) 463-7266

fax: (989) 463-7076

# ABSTRACT

Cross-identifications for 14,574 intrinsically luminous galactic stars (mostly OB stars) to objects in the 2MASS survey have been determined using a search box of ±0.0015 degrees (±5.4 arcsec) in both RA and Dec. Instructions on obtaining the relevant files can be obtained at othello.alma.edu/~reed/OB-2MASS.doc.

## 1  Introduction

The last several years have witnessed significant developments in the application of near-infrared techniques to intrinsically hot galactic stars. Hanson and her collaborators, for example, (Hanson et al. 2005 and references therein) give a brief chronology of these developments and show how analysis of intermediate-resolution, high signal-to-noise H-and K-band spectroscopy of OB stars can be used to constrain the temperature, luminosity, and wind parameters of these objects when used in conjunction with atmosphere-model codes. While one might not normally think of studying OB stars at infrared wavelengths, there is advantage in doing so in view of the significantly lower extinction at these wavelengths as compared with traditional UBV photometry and visual-band spectroscopy. To appreciate this, consider a star of surface temperature 30,000K, about that of a B0 main-sequence object. The Planck distribution indicates that such a star will be about 3.3 magnitudes brighter in the V-band (0.55μ) than at the J-band (1.25μ). (For T → ∞, this figure approaches a limiting value of ~ 3.6 magnitudes.) However, J-band extinction is significantly less than its V-band counterpart: $A_J$ ~ 1.62$E_{J-K}$ vs. $A_V$ ~ 5.82$E_{J-K}$ where $E_{J-K}$ is the J-K color excess, that is, $A_J$ ~ 0.28$A_V$ (Tokunaga 2000). The observed color of such a star will then behave as (J-V) ~ 3.3 – 0.72$A_V$, and for visual extinction $A_V$ greater than about 4.5 magnitudes such a star will actually appear brighter in J than in V. With visual extinction in the galactic plane of typically 1.5 mag/kpc, such extinctions will be the case for stars more distant than about 3 kpc – less than halfway to the galactic center from the Sun. The infrared thus opens a window to distant, highly-reddened OB stars otherwise inaccessible at visual wavelengths. The drawback of IR wavelengths is of course that IR photometric colors alone cannot be used to discern the presence of an OB star: the colors of such stars are essentially degenerate in the infrared and the reddening line is essentially parallel to the main sequence; there is no quantity analogous to the reddening-free Q-parameter of UBV photometry.

In view of the growth of interest in applying IR techniques to hot stars, it was thought that a cross-reference between known galactic OB stars and sources cataloged in the synoptic Two-Micron All-Sky Survey (2MASS; Skrutskie et al. 2006) might be of use to this component of the research community. This is the purpose of this paper.

## 2. Searching the 2MASS Catalog for Galactic OB Stars

This author maintains an all-sky catalog and databases of published UBVβ photometry and MK spectral classifications for intrinsically luminous galactic stars (Reed 2003). At this writing, this catalog includes just over 18,600 individual objects. The vast majority of these are OB stars, along with some giant, bright giant, and supergiant A-G stars. Instructions on accessing the catalog, data files, and supporting documentation are available at othello.alma.edu/~reed/OBfiles.doc. "OB stars" here is taken to mean (O3-B2) V and (O3-B9) I-IV stars. Stars in these files are listed according as an arbitrarily-assigned "Alma Luminous Star" (ALS) number, but the main catalog includes cross-references to source names, HD, HR, SAO, CP, CPD, and BD numbers as well as coordinates. This paper reports development of a file of cross-identifications between ALS and 2MASS objects.

I obtained a copy of the 2MASS survey on DVDs directly from the Infrared Processing and Analysis Center. The 2MASS point-source catalog (PSC) comprises 92 compressed files listing a total of nearly 471 million sources. Each source object (mostly galactic stars) occupies one line of data of 60 fields giving positional information, $JHK_s$ magnitudes and corresponding uncertainty estimates, various quality flags, and cross-reference information to the Tycho-2 catalog. A short program was written to read through each uncompressed file and extract data from the first 28 fields, the most important ones for photometric and quality information. Another program read through these extracted files, discarding any lines where any fields contained null (N) data or where the photometric quality and contamination flags failed to satisfy one or more of the following constraints: photometric quality = AAA, blend flag = 111, contamination flag = 000, galactic contamination flag = 0, and minor planet flag = 0. These constraints are probably on the severe side, but it was thought desirable to err on the side of being conservative with potential cross-identifications. After application of these constraints, some 160 million sources remained.

The search for cross-identifications was carried out by comparing 2MASS equatorial coordinates against decimalized coordinates from the author's OB-star catalog. A search rectangle of half-width 0.0015 degrees (= 5.4 arcsec) was used in both right ascension and declination. This resulted in just over 14,600 hits, including about two dozen cases (mostly objects in crowded fields) where a given OB star was identified with two separate 2MASS sources. A few of these could be resolved manually, but the majority were removed from the final list in order to avoid incorporating suspect identifications. After deletion of such cases, 14,574 cross-references remained.

Instructions on obtaining the listing of cross-references can be obtained at othello.alma.edu/~reed/OB-2MASS.doc. The file itself (OB-2MASS.dat) lists ALS numbers, ALS-catalog source names, 2MASS identifiers, the name of the PSC file in which the 2MASS source appears, and the default 2MASS $JHK_s$ magnitudes and their uncertainties. A fuller version of this paper including the figures described below is also available at this site; the figures are too large to include with the arXiv version of the paper.

### 3. Magnitude and Color Comparisons

Figure 1 shows the IR two-color diagram for the 14,574 2MASS objects identified as luminous galactic (mostly OB) stars; the arrow shows a reddening line with $E_{J-H} \sim 1.7 E_{H-K}$ (Tokunaga 2000). $H-K_S$ color excesses of up to about 1.5 magnitudes are evident, corresponding to $A_V \sim 15.6 E_{H-K} \sim 23$ magnitudes. Figure 2 shows a comparison of 2MASS J and Johnson V magnitudes for 10,414 of these stars for which these data are available in common. As anticipated from the extinction argument given in the Introduction above, the vast majority of these objects have J < V. The small number of stars with J-V > 3.5 are likely misidentifications caused by OB stars with slightly erroneous coordinates; such extreme colors cannot be expected on any physical grounds. These same few scattered points also show up in Figure 3, where J-V is plotted as a function of E(B-V) for 5491 stars; the objects plotted here are restricted to types (O3-B2) V and (O3-B9) I-IV. The bulk of the points in this figure lie along a line of slope $\sim -2.2$, very close to what one would expect for $A_V \sim 3.2 E_{B-V}$.